\documentclass[10pt,jorunal]{IEEEtran}

\usepackage{fst}

\usepackage{amsmath,amssymb}
\usepackage[colorlinks=true,bookmarks=false,citecolor=blue,urlcolor=blue]{hyperref} %

\newcommand{\hl}[1]{#1}
\newcommand{\hll}[1]{#1}

\usetikzlibrary{external}

\newcommand{\pltref}[2]{\tikzset{external/export next=false}\tikz[baseline=-0.55ex]{\draw[thick, #1] (0,0) -- (0.6,0) plot[line width=1,mark options={solid,scale=0.5},mark=#2] coordinates{(0.3,0)};}} %

\tikzsetexternalprefix{figures_compiled/}
\begin{document}

\title{Probabilistically Shaped 4-PAM for Short-Reach IM/DD Links with a Peak Power Constraint}

\author{Thomas Wiegart,~\IEEEmembership{Student~Member,~IEEE}, Francesco Da Ros,~\IEEEmembership{Senior Member,~IEEE}, Metodi Plamenov Yankov,~\IEEEmembership{Member,~IEEE}, Fabian Steiner,~\IEEEmembership{Student~Member,~IEEE}, Simone Gaiarin,~\IEEEmembership{Member,~IEEE},\\Richard D. Wesel,~\IEEEmembership{Senior~Member,~IEEE}%
 \thanks{T.~Wiegart and F.~Steiner are with the Institute for Communications Engineering, Technical University of Munich (TUM). E-Mail(s): \texttt{\{thomas.wiegart, fabian.steiner\}@tum.de}}%
 \thanks{F.~Da Ros, M.~P.~Yankov, and S.~Gaiarin are with DTU Fotonik, Technical University of Denmark (DTU). E-Mail(s): \texttt{\{fdro,meya,simga\}@fotonik.dtu.dk}} %
 \thanks{ R.~Wesel is with the Communications System Laboratory, University of California Los Angeles (UCLA).}
  \thanks{E-Mail: \texttt{wesel@ucla.edu}.}%
}

\maketitle

\vspace{-0.2cm}
\begin{abstract}
Probabilistic shaping for intensity modulation and direct detection (IM/DD) links is discussed and a peak power constraint determined by the limited modulation extinction ratio (ER) of optical modulators is introduced. The input distribution of 4-ary unipolar pulse amplitude modulation (PAM) symbols is optimized for short-reach transmission links without optical amplification nor in-line dispersion compensation. 
The resulting distribution is symmetric around its mean allowing to use probabilistic amplitude shaping (PAS) to generate symbols that are protected by forward error correction (FEC) and that have the optimal input distribution. The numerical analysis is confirmed experimentally for both an additive white Gaussian noise (AWGN) channel and a fiber channel, showing gains in transmission reach and transmission rate, as well as rate adaptability.
\end{abstract}


\section{Introduction}

Noncoherent transceivers based on intensity modulation (IM)
and direct detection (DD) are currently the preferred choice for short reach links, e.g., data center interconnects or
metro connections up to \SI{100}{km}. Here the key requirements are low power consumption, low complexity, low cost, and minimum latency~\cite{zhong_shortreach_review}.
As a result, short reach links focus on avoiding energy-hungry optical amplifiers and in-line optical dispersion compensation 
\hll{as these components would increase the latency and the power budget of the system.}

Recently, probabilistic shaping (PS) and especially probabilistic amplitude shaping 
(PAS)~\cite{bocherer_bandwidth_2015,buchali_rate_2016} received broad interest from 
academia and industry for coherent optical communications. PAS enables rate adaptation and a performance close to the Shannon limit. PAS 
is compatible with standard forward-error correction (FEC) schemes, but requires a
symmetric input distribution to synthesize the optimal distribution after FEC encoding.
Most works apply PS to an average power constrained additive white Gaussian (AWGN) channel
with quadrature-amplitude modulation (QAM). 
Only recently, PS has started  being investigated for IM/DD transmission, which imposes a non-negativity constraint on the channel input $X$, as it is used to modulate the intensity of the electric field. In this case, an average (optical) power constraint is expressed
as $\E{X}$ \cite{kahn_wireless_1997}, and the resulting optimal input distribution is asymmetric (e.g., \cite{eriksson_56_2017}) so that PAS can not be employed directly.

In \cite{eriksson_56_2017}, the authors considered 4-ary
pulse-amplitude modulation (PAM), optimized the input distribution and analyzed achievable rates for
\SI{56}{GBaud} transmission. %
This first experimental demonstration included both optical amplification and in-line dispersion compensation. Under these conditions, an average-power constraint was applied in the PS optimization, resulting in an exponential distribution that yields close to optimal performance. However, no coded results are provided. %
By using optical amplification, the system was not limited by the ER of the optical modulator, which normally limits the launch power in practical transmission scenarios, thus justifying the average-power constraint. PS for similar setups including optical amplifiers has recently also been considered in, e.g., \cite{he_ldpc2020, zhang_pspam8_2020}.

 In \cite{he_probabilistically_2019} and \cite{kim2020transmission}, instead, the authors consider 4-PAM and 8-PAM for un-amplified transmission. They propose to slightly modify the exponential distribution and assign the same probability to two symbols such that using PAS for FEC integration is possible. Improvements in receiver sensitivity of approximately $\SI{1}{dB}$ are observed. However, the additional modulator loss for the PS modulation format is not taken into account.
As the authors point out in \cite{he_probabilistically_2019}, operating the PS modulation format with the same ER as the uniform modulation format leads to a reduction in launch power by $\SI{1.5}{dB}$.
Overall, the reduction of launch power is higher than the gain in receiver sensitivity and thus PS with an exponential-like distribution effectively shortens the reach.
The reduced launch power
for the exponential-like distribution optimized with an average power constraint can not be compensated in the absence of an optical amplifier.

In this work, we investigate PS for practical IM/DD links with no optical amplifiers nor optical dispersion compensation. Unlike previous works, we introduce  a peak power constraint in the PS optimization.
The peak-power constraint originates from the need to avoid increasing the optical loss in the Mach Zehnder modulator (MZM) due to its limited ER. The same peak-to-peak driving voltage is considered for both uniform and PS formats and similar optical loss are therefore targeted to avoid penalizing the PS modulation. This approach is even more relevant when alternative modulation schemes are considered, e.g., directly modulated lasers or electro-absorption modulated lasers~\cite{zhang_IMDD_comparison}. 

The peak power constraint implies that the optimal distribution is discrete with a finite number of mass points~\cite{smith_information_1971}.
The optimization of the achievable rate for symbol-metric decoding (SMD)
and bit-metric decoding (BMD) shows that the optimal distribution 
is symmetric around $\E{X}$ and is fundamentally different from the ones suggested previously (e.g., \cite{he_probabilistically_2019}). 
We discuss potential gains using a linear AWGN model first and conduct optical back-to-back (B2B), i.e., AWGN channel, and realistic transmission experiments for 4-PAM transmission with symbol rates between $\SI{8}{GBaud}$ and $\SI{32}{GBaud}$.

The remaining of the paper is organized as follows. In Section~\ref{sec:system_model} the model of the extinction ratio constrained IM/DD system is introduced, and the PS optimization is discussed showing achievable bit- and symbol- metric decoding performance. In Section~\ref{sec:experiment} the setup used for the experimental validation is described and the measurement results are reported. In Section~\ref{sec:conclusion} the main conclusions are drawn.

\section{System Model}
\label{sec:system_model}

The IM/DD link is modeled by
\begin{equation}
    Y = \abs{ \sqrt{\abs{X}} + N_\text{opt} }^2 + N_\text{el}
\end{equation}
where $N_\text{opt}$ and $N_\text{el}$ denote additive noise in the optical and electrical domain, respectively, and the random variable (RV) $X$ denoting the transmit symbol has alphabet
\begin{equation}\cX = \{0, 1, 2, 3\}.\end{equation}
As $X$ is modulated onto the intensity (i.e., the power) of the optical signal, the optical amplitude is proportional to the square root of the absolute value of $X$. At the receiver, square-law detection is performed, i.e., one obtains an electrical current proportional to the squared absolute value of the impinging optical amplitude. Due to typically low transmit powers of IM/DD systems, the Kerr nonlinearity of the fiber was neglected. This model also does not capture the impact of chromatic dispersion (CD). We will discuss the validity of this model based on the experimental results in Sec.~\ref{sec:experiment}.

In this paper, we consider un-amplified links and the signal laser has an optical SNR in excess of $\SI{45}{dB}$. The noise contributions are thus dominated by electrical noise sources, e.g., electrical transmitter and receiver amplifiers, digital-to-analog and analog-to-digital converter noise. By neglecting the optical noise contributions, the system model simplifies to a linear AWGN model of the form 
\begin{align}
    Y = \label{eq:model} |X| + N_\text{el}
\end{align}
where $N_\text{el}$ is a Gaussian distributed RV with zero mean and variance $\sigma^2$.  In order to ensure a fair comparison, we aim for the same ER for uniform and shaped signaling, i.e., the MZM is operated with the same peak-to-peak driving voltage for both modes. Thus we have a peak power constraint on the input signal $X$, which is implicitly included in the model by constraining $X$ to the set $\cX$.
We define the  peak-signal-to-noise ratio (PSNR) in the electrical domain as
\begin{equation}
    \operatorname{PSNR} = \frac{\max\{\cX\}^2}{\sigma^2} = \frac{3^2}{\sigma^2}.
\end{equation}

\begin{figure}[t]
    \centering
    \footnotesize
    \includegraphics{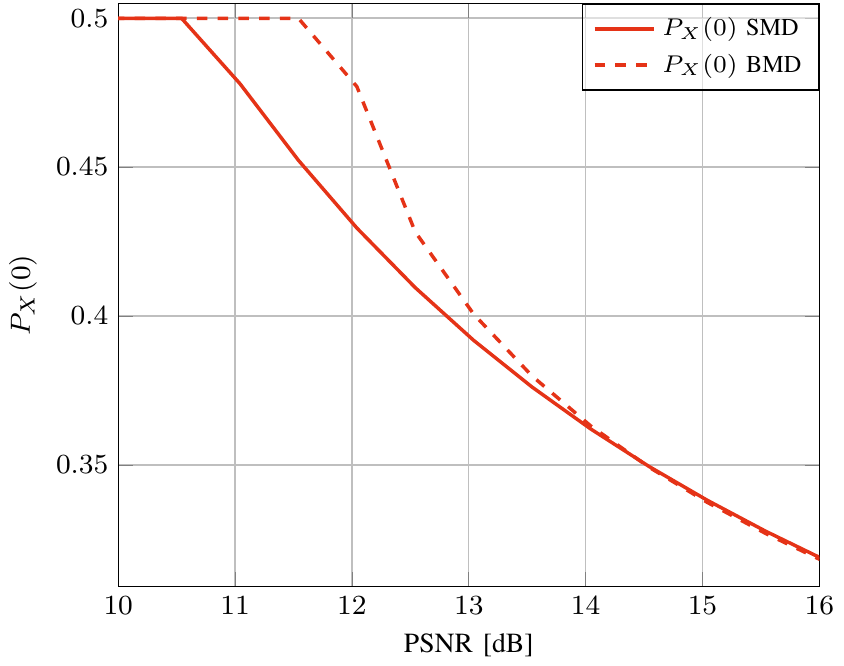}
    \caption{Optimal $P_X(0)$ for SMD and BMD. If $P_X(0) = 0.5$, this corresponds to OOK signalling, whereas $P_X(0) < 0.5$ is $4$-PAM signalling.}
    \label{fig:optimal_pX0}
\end{figure}

Depending on the FEC architecture, either SMD or BMD metrics are relevant~\cite{bocherer_probabilistic_2019-1}.
For BMD we label each constellation point $x\in\cX$ with a two bit binary label, 
$x \mapsto \vb = (b_1b_2)$. A binary
reflected Gray code \cite{gray1953pulse} performs well. To find the optimal distribution for PS, we solve the
problem
\begin{align}
    \max_{P_X} \quad R_{\tsmd/\tbmd} \qquad \text{subject to}\quad X \in \cX\label{eq:optim_problem}
    \vspace{-0.2cm}
\end{align}
where 
\begin{equation} R_\tsmd = \I(X;Y) = \entr(X) - \entr(X|Y) \label{eq:Rsmd} \end{equation}
and 
\begin{equation}R_\tbmd = \left[\entr(X) - \sum_{k=1}^2 \entr(B_k|Y)\right]^+ \label{eq:Rbmd} \end{equation}
\begin{figure*}[t]
    \centering
    \subfloat[Achievable rates with BMD.]{
        \footnotesize
        \includegraphics{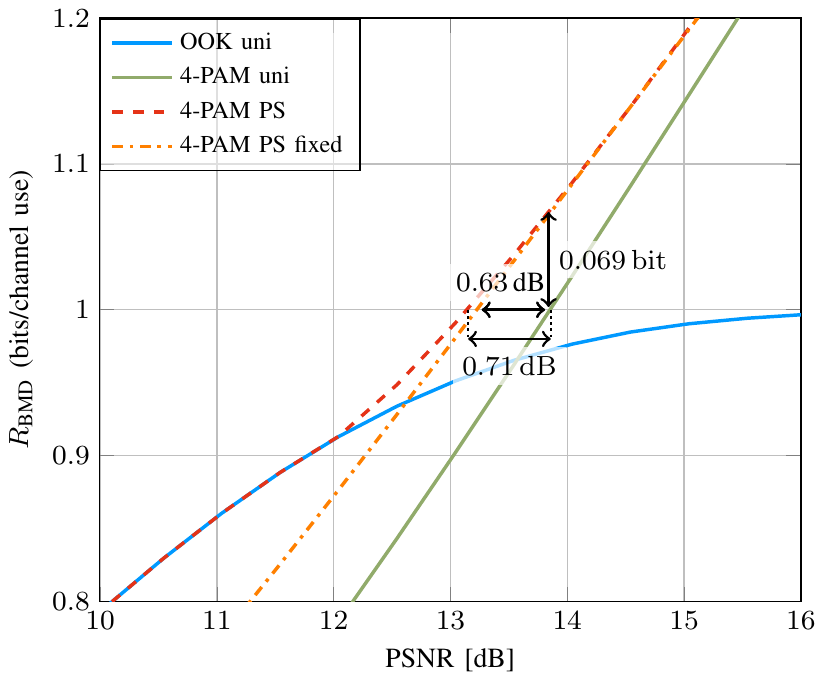}
        \label{fig:rates_linear_model_bmd}
    }
    \hfill
    \subfloat[Achievable rates with SMD.]{
        \footnotesize
         \includegraphics{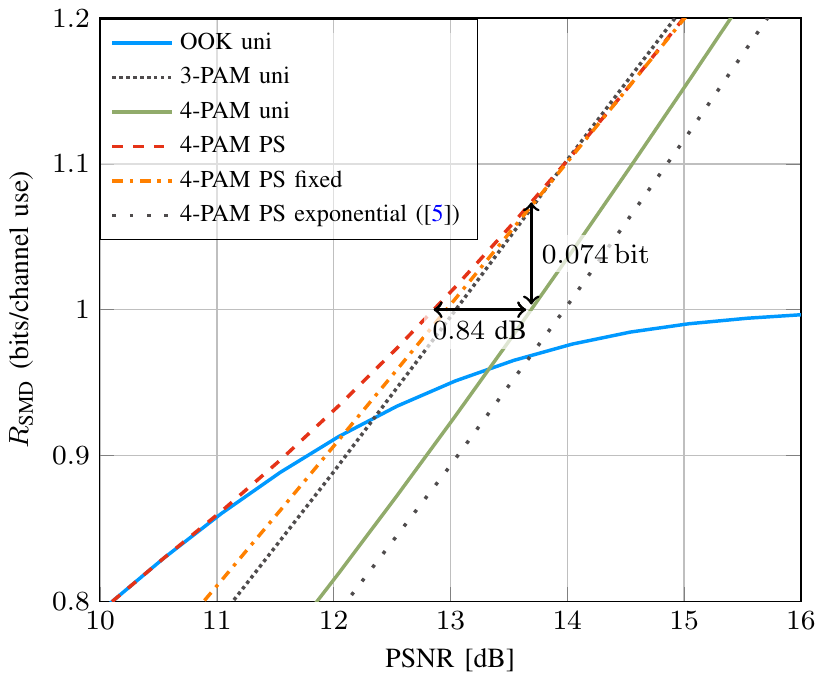}
        \label{fig:rates_linear_model_smd}
    }
    \caption{Achievable rates assuming the linear system model \eqref{eq:model} and using BMD (a) and SMD (b), respectively. \hll{The distribution for the shaped curve (4-PAM PS) is chosen according to Fig.~\ref{fig:optimal_pX0}.}}
    \label{fig:rates_linear_model}
\end{figure*}
\begin{figure}[t]
\centering
    \footnotesize
     \includegraphics{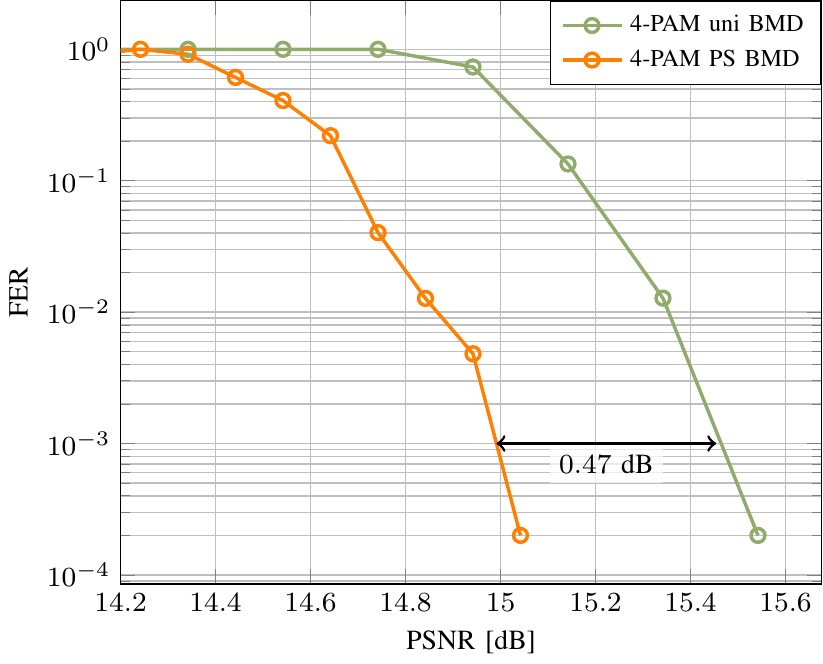}
    \caption{Coded results for 4-PAM \hll{at a spectral efficiency of $\SI{1}{bpcu}$} assuming the linear system model.}
    \label{fig:coded_linear_model}
\end{figure}
with $[\cdot]^+ = \max(0,\cdot)$ and where $\I(\cdot;\cdot)$ denotes mutual information (MI), $\entr(\cdot)$ denotes entropy and $\entr(\cdot|\cdot)$ denotes conditional entropy. 
The constraint $x \in \cX$ is equivalent to a peak power constraint and a uniform spacing constraint on the signaling points. Using a similar approach as in
\cite[Proposition~2.3]{huang_characterization_2005} and exploiting the fact that the mutual information is invariant with regard to a shift in the input constellation,
it can be shown that the optimal distribution is symmetric around its mean $\E{X}$. 
\hl{Intuitively, this is the case because the input $X$ has an amplitude constraint on both sides (due to the peak power constraint and the non-negativity constraint) and the channel is linear.}
For 4-PAM and fixed mass point locations in $\cX$, the optimization \eqref{eq:optim_problem}
boils down to a linesearch over a single parameter, namely the probability $P_X(0) \in [0,0.5]$, as $P_X(3) = P_X(0)$ and $P_X(1) = P_X(2) = (1-2\cdot P_X(0))/2$.

The optimal $P_X(0)$ is shown in Fig.~\ref{fig:optimal_pX0} for SMD and BMD. The plot also indicates the optimal constellation cardinality for each noise power. For low noise powers, $P_X(0) = \num{0.5}$ indicates that the optimal signaling uses a constellation with two mass points only, i.e., OOK. The optimal constellation cardinality depends on the decoding metric: for BMD, the operation range of OOK is larger than for SMD. \hll{With increasing PSNR the optimal distribution changes from 4-PAM with high probability mass at the outer two points towards uniform 4-PAM.}

The corresponding achievable rates for uniform \hll{(``4-PAM uni'' in the legend)} and optimized PS signaling \hll{(``4-PAM PS'')} are shown in Figs.~\ref{fig:rates_linear_model_bmd} and \ref{fig:rates_linear_model_smd} for BMD and SMD, respectively.
We observe a gain of \SI{0.83}{dB} in SNR for PS with SMD at \SI{1.0}{bits/channel use (\bpcu)}. For BMD, the gain is \SI{0.71}{dB}. We also depict the achievable rates
for the distribution $P_X = (0.35, 0.15, 0.15, 0.35)$ (``4-PAM PS fixed") which will be used in the optical experiment of Sec.~\ref{sec:experiment}. As can be inferred from Fig.~\ref{fig:rates_linear_model}, \hll{this $P_X$ performs close to the optimal distribution for a wide range of operating points.}
Here, the gain for BMD is \SI{0.63}{dB} at a rate of $\SI{1.0}{bpcu}$. Compared to the bipolar setting (see~\cite[Table~3]{bocherer_bandwidth_2015}), BMD with PS exhibits a non-negligible gap to SMD for the unipolar case. The gap between SMD and BMD can be closed by using a non-binary FEC code or using a binary FEC code with multi-level coding and multi-stage decoding \cite{wachsmann1999multilevel}.

\begin{figure*}[t]
    \centering
    \footnotesize
    \subfloat[Setup for back to back and transmission experiments.] {
        \includegraphics[scale=0.8]{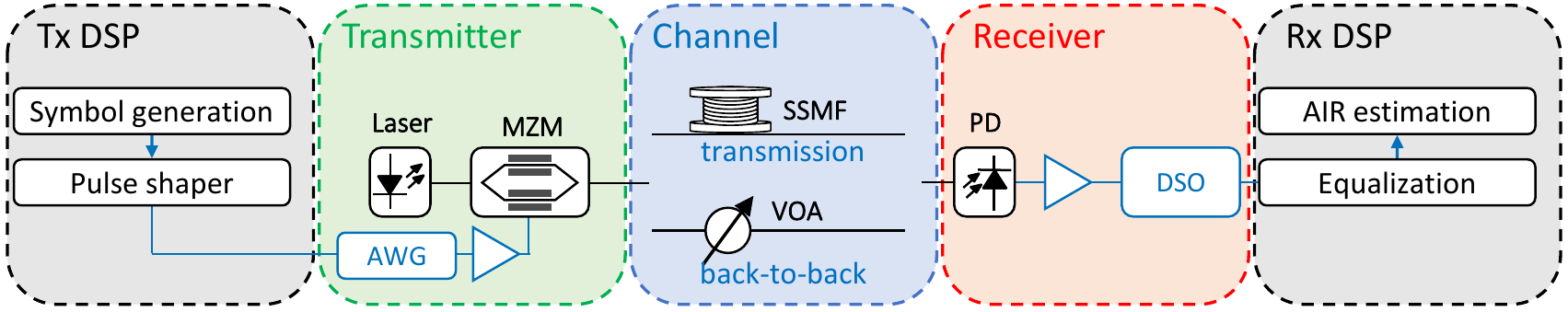}
        \label{fig:experimental_setup}
    }
    \hfill
    \subfloat[Input distribution used for the experimental setup.]{
        \footnotesize
        \includegraphics{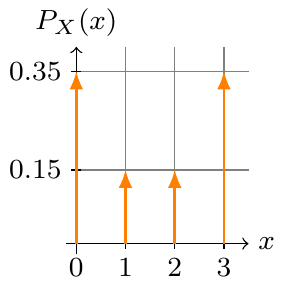}
        \label{fig:distribution}
    }
    \caption{Experimental setup and optimized input distribution.}
\end{figure*}
\begin{figure*}[t]
\centering
\fbox{\parbox{.95\linewidth}{\footnotesize{\hfill
\pltref{TUMBeamerRed}{*}/\pltref{TUMBeamerRed,dashed}{*} $\SI{10}{GBaud}$ uni/shp\hfill
\pltref{TUMBeamerOrange}{*}/\pltref{TUMBeamerOrange,dashed}{*} $\SI{16}{GBaud}$ uni/shp\hfill
\pltref{TUMBeamerBlue}{*}/\pltref{TUMBeamerBlue,dashed}{*} $\SI{20}{GBaud}$ uni/shp\hfill
\pltref{TUMBeamerGreen}{*}/\pltref{TUMBeamerGreen,dashed}{*} $\SI{32}{GBaud}$ uni/shp \hfill{\phantom{.}}} %
}
}\vspace{-0.2cm}

    \subfloat[$\hat R_\tbmd$ estimates versus optical attenuation of the VOA.]{
        \footnotesize
         \includegraphics{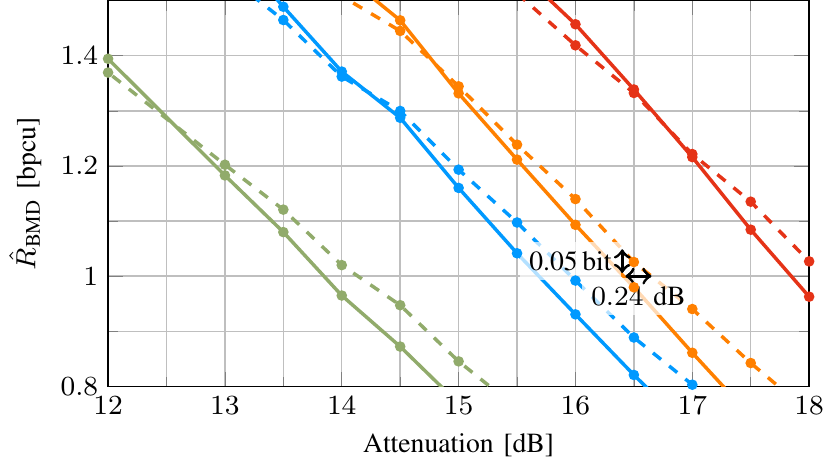}
        \label{fig:bit_b2b_results}
    }
    \hfil
    \subfloat[$\hat R_\tsmd$ estimates versus optical attenuation of the VOA.]{
        \footnotesize
         \includegraphics{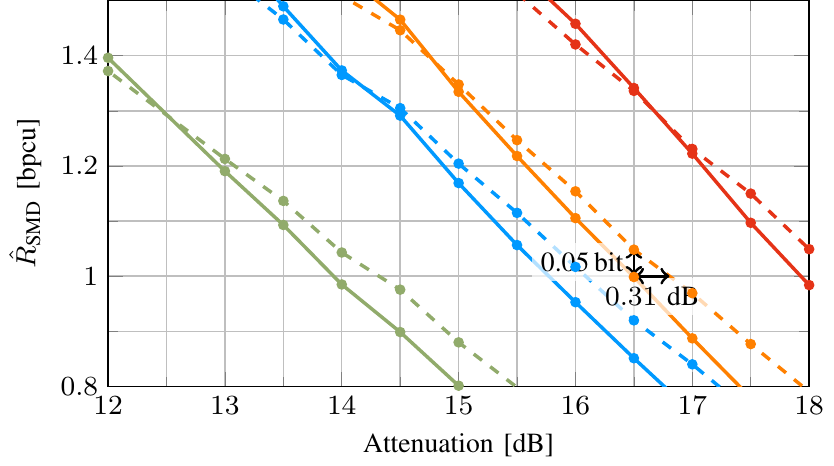}
        \label{fig:symb_b2b_results}
    }
    \hfill
    
    \subfloat[$\hat R_\tbmd$ estimates versus fiber length for transmission over SSMF.]{
        \footnotesize
         \includegraphics{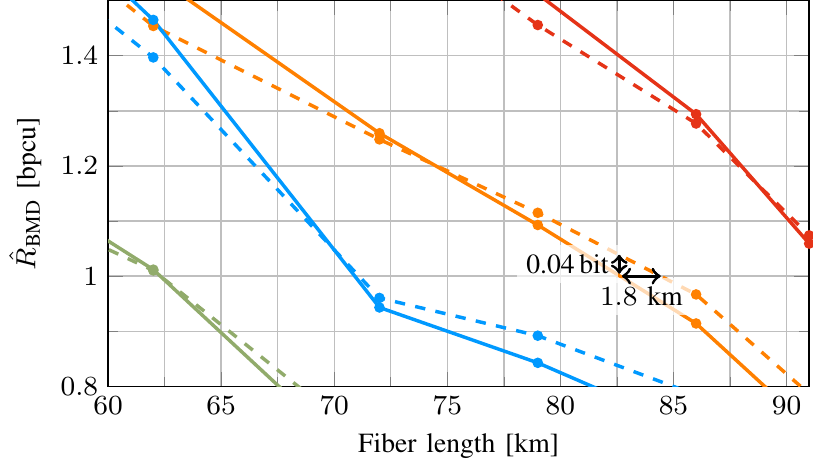}
        \label{fig:bit_fiber_results}
    }
    \hfill
    \subfloat[$\hat R_\tsmd$ estimates versus fiber length for transmission over SSMF.]{
        \footnotesize
         \includegraphics{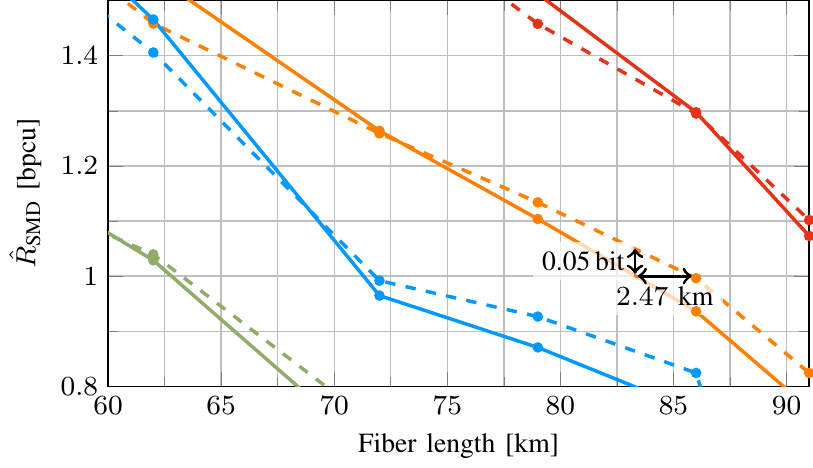}
        \label{fig:symb_fiber_results}
    }

\caption{Experimental results for uniform and shaped (with the distribution depicted in Fig.~\ref{fig:distribution}) transmission and different symbol rates: (a) and (b) depict $\hat R_\tbmd$ and $\hat R_\tsmd$ estimates with linear optical attenuation, respectively.  and (c) and (d) depict $\hat R_\tbmd$ and $\hat R_\tsmd$ estimates for fiber transmission including dispersion effects, respectively.}
\label{fig:experiment_results}
\end{figure*}

\begin{figure*}
    \footnotesize
    \centering
     \includegraphics{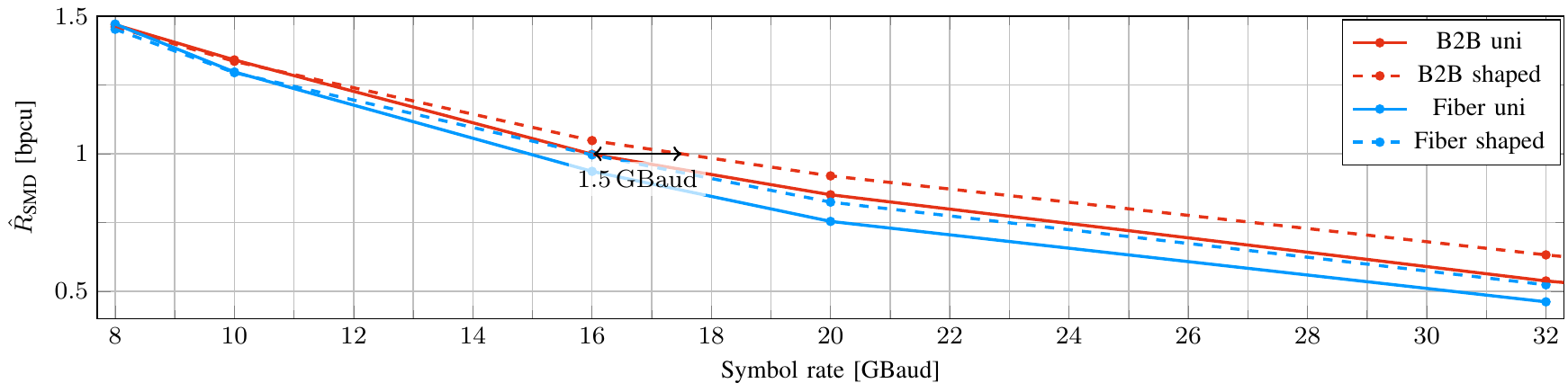}
    \caption{$\hat R_\tsmd$ estimates for uniform (solid curves) and shaped (dashed curves) transmission versus symbol rate for back-to-back measurements with optical attenuation of the VOA of $\SI{16.5}{dB}$ (red curves) and for fiber transmission with fiber length $L=\SI{86}{km}$ (blue curves).}
    \label{fig:symb_fiber_rate}
\end{figure*}

\begin{figure}
    \footnotesize
    \centering
     \includegraphics{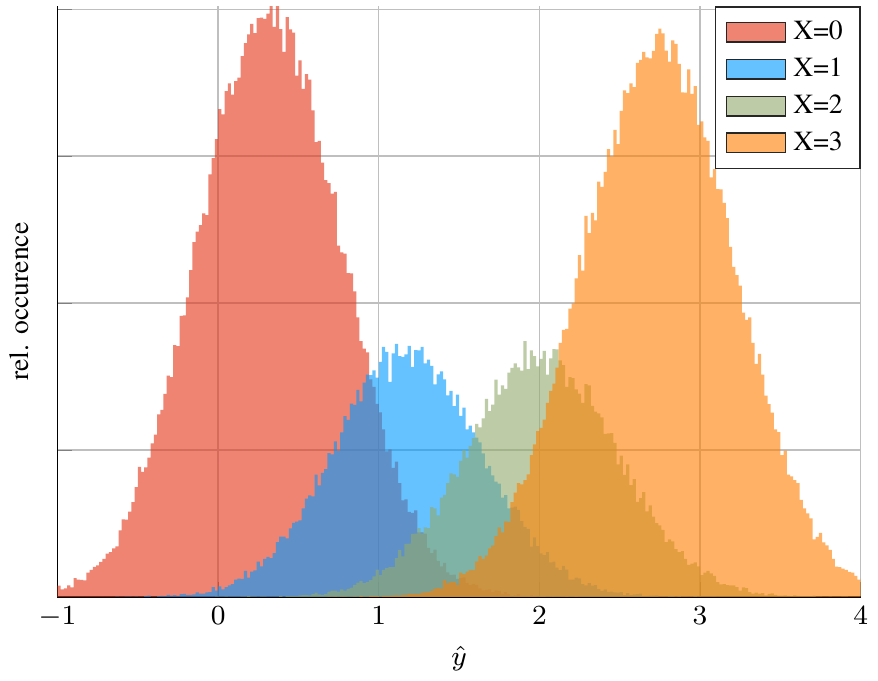}
    \caption{\hll{Histogram of the estimated receive symbols $\hat{y}$ after the equalizer for transmit symbols $X \in \cX$ with the distribution from Fig.~\ref{fig:distribution}, transmitted with a symbol rate of $\SI{16}{Gbaud}$ over a fiber distance of $L=\SI{79}{km}$.}}
    \label{fig:hist}
\end{figure}

As the optimal input distribution $P_X$ is symmetric around $\E{X}$, we can use PAS with FEC. This is in contrast to previous works \cite{eriksson_56_2017,he_probabilistically_2019} where an average power constraint is considered and the best distribution is asymmetric. 

The exponential distribution from~\cite{eriksson_56_2017}, optimized for an average power constraint, is also shown in Fig.~\ref{fig:rates_linear_model_smd} as reference (``4-PAM PS exponential''). \hll{In the depicted range of Fig.~\ref{fig:rates_linear_model_smd},} the achievable rate for such a distribution under a realistic peak power constraint is even lower than in the uniform case.

We validate the information theoretic results of Fig.~\ref{fig:rates_linear_model} through finite length simulations with low-density parity-check (LDPC) codes in Fig.~\ref{fig:coded_linear_model}. The LDPC code \hl{is regular and} has degree 3 variable nodes (VNs) and a blocklength of around $10^4$ bits. \hll{The code was constructed using a protograph based design} \cite{divsalar_capacity_2009} \hll{with girth optimization techniques ensuring a girth of at least six.} 
The uniform scenario uses a rate 1/2 code (100\% OH), while the PS scenario uses a rate 0.56 (78\% OH) code with PAS and \ac{CCDM} \cite{schulte_constant_2016}. We use sum-product decoding with 50 iterations. At a frame error rate (FER) of \num{e-3}, we observe a gain of \SI{0.47}{dB}. Tailored code designs, e.g.,~\cite{steiner_protograph-based_2016}, can further improve the performance.

Although the achievable gains in terms of power efficiency are smaller in a peak-power limited non-coherent setting compared to average-power limited coherent transmission, we emphasize that probabilistic shaping can also be used to introduce rate adaptivity to the transmission system, by adjusting the shaping and FEC overheads. In particular, one can operate 4-PAM at PSNR levels where (uniform) 3-PAM (see black curve in Fig.~\ref{fig:rates_linear_model_smd}) would be used. By operating with 4-PAM instead of 3-PAM, an easier integration with binary FEC and binary input data is possible.

\section{Experimental Setup and Results}
\label{sec:experiment}

The experimental setup is shown in Fig.~\ref{fig:experimental_setup}.
Uniform or shaped (with the distribution shown in Fig.~\ref{fig:distribution}) 4-PAM sequences are generated, rectangular pulse shaping is applied and the time domain waveforms are loaded into an arbitrary waveform generator (AWG, 64 GSa/symbol). The AWG output is amplified and used to drive a MZM in order to encode the 4-PAM signal onto an optical carrier from an external cavity laser at 1550 nm. The peak-to-peak driving voltage has been kept constant for all constellation shapes considered and was chosen to operate the MZM in the linear regime. Two scenarios have been investigated, a B2B configuration where the received power was varied via a variable optical attenuator (VOA) and a transmission scenario where different lengths of SSMF ($\alpha = \SI{0.2}{dB/km}$, $D = \SI{17}{ps/nm\cdot km}$) were considered. The launch power into the SSMF was $\SI{2.5}{dBm}$, a limitation imposed by the insertion loss of the MZM. The baudrate was varied between $\SI{8}{GBaud}$ and $\SI{32}{GBaud}$. At the receiver, a $\SI{45}{GHz}$ photodetector (PD) followed by a low-noise electrical amplifier converts the signal back to the electrical domain before analog-to-digital (A/D) conversion using a 80-GSa/s digital storage oscilloscope (DSO). Offline digital signal processing (DSP) is then applied, consisting of 
a linear feed forward equalizer (FFE, 5 taps for B2B, 11 taps for fiber transmission), to partially mitigate for the impact of chromatic dispersion and frequency limitations of electrical components, and achievable information rate (AIR) estimation~\cite[Sec.~VII-G]{bocherer_probabilistic_2019-1}.
The estimated AIR with SMD and BMD is denoted by $\hat R_\tsmd$ and $\hat R_\tbmd$, respectively.

\hll{
From the definition of the AIR in \eqref{eq:Rsmd} and \eqref{eq:Rbmd} for SMD and BMD, respectively, it can be seen that the AIR contains both required overheads for shaping and forward error correction and thus provides a fair comparison between uniform and shaped signalling \cite{bocherer_probabilistic_2019-1}. 
}

We depict the B2B results for BMD and SMD in Figs.~\ref{fig:bit_b2b_results} and \ref{fig:symb_b2b_results}, respectively. The results considering this linear channel are in good agreement with the performance calculated for the channel model of (\ref{eq:model}) shown in Fig.~\ref{fig:rates_linear_model}. Considering a symbol rate of $\SI{16}{GBaud}$ and for a target achievable rate of \SI{1}{\bpcu}, we observe that PS signaling allows an optical attenuation which is \SI{0.24}{dB} and  \SI{0.31}{dB} higher than for the uniform case, for BMD and SMD, respectively. Alternatively,  the rate gain is approximately \SI{0.05}{\bpcu} ($\approx 5$\%) and such a gain is consistent across the symbol rates investigated.

The transmission results over SSMF of different length are shown in Figs.~\ref{fig:bit_fiber_results} and~\ref{fig:symb_fiber_results}, again for BMD and SMD respectively. As the transmission length was varied by combining a different number of fiber spools of different lengths, the optical loss did not scale linearly with the transmission distance. Therefore, even though the CD scales linearly with distance, the rate decrease deviates from being monothonical with the increased fiber length. Nevertheless, for a symbol rate of $\SI{16}{GBaud}$, we observe a rate gain of $\SI{0.05}{bpcu}$ ($\approx 5$\%), i.e., \SI{0.8}{Gbit/s}, compared to uniform signaling or a reach gain of up to $\SI{2.4}{km}$  ($\approx 3$\% improvement) for SMD. The improvement slightly decreases down to 
$\SI{1.8}{km}$ for BMD. %
These results are still in line with those from the B2B setting. The key difference with the B2B results (Figs.~\ref{fig:bit_b2b_results} and \ref{fig:symb_b2b_results}), is the impact of CD which becomes more significant as the symbol rate increases, reducing the improvement provided by shaping (at the a target achievable rate of \SI{1}{\bpcu}) for higher symbol rates. The impact of dispersion could be decreased by considering more complex equalization schemes which are beyond the scope of this work. More interesting, CD could be included within the channel model used for the shaping optimization. However, that would require to consider temporal effects within the shaping (i.e., shaping across several symbols), yielding a steep increase in complexity, as shown for coherent systems~\cite{yankov2017temporal}.
\hl{In the experiment, we observed that a slightly more uniform distribution shows better performance when the signal is affected by CD: the distribution depicted in Fig.~\ref{fig:distribution} with $P_X(0) = 0.35$ achieves slightly higher rates than the one optimized for the dispersion-less channel model of~\eqref{eq:model} ($P_X(0) = 0.4$ for a spectral efficiency of $\SI{1}{bpcu}$).}

\hll{Fig.~\ref{fig:hist} depicts the histogram of the estimated receive symbols $\hat{y}$ after the equalizer for transmission over $\SI{79}{km}$ of SSMF with a symbol rate of $\SI{16}{Gbaud}$. The transmit symbols were chosen with the distribution in Fig.~\ref{fig:distribution} and we distinguish between the transmit symbols in $\cX$ by different colors. Thus, Fig.~\ref{fig:hist} shows the empirical observation of the end-to-end channel law $p_{Y|X}$. This confirms the validity of the assumed linear channel model \eqref{eq:model} with additive Gaussian noise.
}

Finally, Fig.~\ref{fig:symb_fiber_rate} shows the rate achievable with SMD as a function of the transmitter symbol rate for a fiber distance of $\SI{86}{km}$ and for an optical attenuation of $\SI{16.5}{dB}$. At the target rate of $\SI{1}{bpcu}$, both B2B and fiber channel lead to an increase in transmission rate of approx. $\SI{1.5}{GBaud}$.

\section{Conclusion}
\label{sec:conclusion}

We investigated PS of 4-PAM for short reach links using IM/DD. In contrast to previous works, we investigated a more realistic short reach transmission scenario without optical amplification nor in-line dispersion compensation. More importantly, our work considers for the first time that the modulated signal is limited by the finite ER of the modulator and thus the peak power of the input symbols is constrained. Interestingly, the optimal input distribution with a peak power constraint has a symmetry and we can directly use PAS. We showed numerical gains of $\SI{0.47}{dB}$ using PAS and LDPC codes. The numerical analysis is validated through optical experiments both for B2B (AWGN channel) and fiber transmission. For a spectral efficiency of $\SI{1}{bpcu}$ with symbol rates between $\SI{8}{GBaud}$ and  $\SI{32}{GBaud}$, we show a reach gain of $2.9\%$ compared to uniform signalling, or, alternatively, a transmission rate gain of $\SI{1.5}{GBaud}$. While the gains for B2B are consistent across the symbol rate range considered, for the fiber channel, the impact of dispersion decreases the shaping gain at higher symbol rates, as expected since CD is not accounted for in our simple channel model.

\section*{Acknowledgements}
This work was supported by the DFG (grants KR 3517/9-1 and KR 3517/8-1), the ERC CoG FRECOM (grant n. 771878), the DNRF Research CoE SPOC (ref. DNRF123), the Villum Young Investigator program OPTIC-AI (grant n. 29344) and the  National Science Foundation (NSF) grant CCF-1911166.  Any opinions, findings, and conclusions or recommendations expressed in this material are those of the authors and do not necessarily reflect the views of the NSF.

\bibliographystyle{IEEEtran}
    \bibliography{IEEEabrv,confs-jrnls,literature2}

\end{document}